\documentclass[conference]{IEEEtran}

\usepackage{cite}
\usepackage{amsmath}
\usepackage{graphicx}
\usepackage{url}
\usepackage[T1]{fontenc}
\usepackage{verbatim}

\title{District-Level Food Environment Indicators and Social Vulnerability in São Paulo}

\author{
\IEEEauthorblockN{Pedro Lemes Sixel Lobo, Eric Tokuda, Kuruvilla Joseph Abraham, Roberto Fray,\\
Dirce Maria Marchioni, Alexandre Cláudio Botazzo Delbem, and Rogerio Salvini \vspace{.3cm}}
\IEEEauthorblockA{Instituto de Informática, Universidade Federal de Goiás, Goiânia, GO, Brazil\\
Instituto de Ciências Matemáticas e de Computação, Universidade de São Paulo, São Carlos, SP, Brazil\\
Escola Superior de Agricultura Luiz de Queiroz da Universidade de São Paulo, USP, Piracicaba, SP, Brazil\\
Faculdade de Saúde Pública, USP, São Paulo, SP, Brazil}
}

\begin{document}

\maketitle
\begin{abstract}
Urban food environments may reflect broader socioeconomic inequalities, but district-level evidence remains limited in Brazilian cities. This study examined whether indicators of food retail and street-market availability discriminate between levels of social vulnerability across the 96 districts of São Paulo. We conducted an exploratory cross-sectional ecological analysis integrating the São Paulo Social Vulnerability Index (IPVS), establishment records from the Relação Anual de Informações Sociais (RAIS), and street-market data from CAISAN. Census-sector information was aggregated at the district level. Twenty districts without an IPVS classification were excluded, resulting in 76 observations. The outcome distinguished districts classified as IPVS level 1 from those classified as levels 2--7. Predictors described the densities of healthy and unhealthy food establishments, the number of street markets, and the availability of establishments selling fresh or \textit{in natura} food. Eight conventional machine-learning classifiers were evaluated using leave-one-out cross-validation. Reported mean F-scores ranged from 0.62 to 0.75, with XGBoost obtaining the highest value. In the Random Forest model, the densities of healthy and unhealthy food establishments jointly accounted for approximately 60\% of the total impurity-based feature importance. These findings indicate that publicly available food-environment indicators contain information associated with the district-level distribution of social vulnerability. However, the small ecological sample, class imbalance, outcome binarization, and cross-sectional design limit predictive generalization and preclude causal or household-level interpretations.
\end{abstract}


\section{Introduction}

Access to affordable and nutritious food is an important component of urban public health. Its distribution, however, is shaped by socioeconomic and spatial inequalities that may leave disadvantaged populations with fewer opportunities to obtain fresh and minimally processed foods. Research on food deserts and other forms of unequal food availability has shown that the characteristics of local food environments frequently differ according to neighborhood socioeconomic conditions \cite{walker2010disparities}. Examining these differences is particularly relevant in large and heterogeneous cities, where aggregate municipal indicators can conceal substantial inequalities among districts.

Urban food environments encompass the location, density, and types of establishments through which residents obtain food. These characteristics may be related to social vulnerability through several mechanisms, including household purchasing power, commercial investment, transportation infrastructure, and the spatial concentration of public and private services. Nevertheless, area-level availability does not directly measure household food consumption or food insecurity. District-level food-environment indicators should therefore be interpreted as contextual characteristics that may be associated with, but are not equivalent to, individual access to food.

The growing availability of administrative and geospatial data has enabled quantitative analyses of relationships between urban environments and socioeconomic or health outcomes. Machine-learning methods have been applied to support the identification of complex patterns relevant to public policy and resource allocation \cite{mcguire2019defining}. Previous studies have used geospatial predictors to examine childhood obesity \cite{ayala2022leveraging}, while satellite imagery and transfer learning have been employed to estimate poverty in settings where conventional socioeconomic data are limited \cite{jean2016combining}. More broadly, large urban datasets have expanded the possibilities for studying inequality at fine geographical scales \cite{glaeser2018big}. Tree-based algorithms, including XGBoost, are commonly used in this context because they can represent nonlinear relationships and interactions among predictors \cite{chen2016xgboost}.

In Brazil, studies of food insecurity and socioeconomic disadvantage have often focused on households and individual-level determinants \cite{Dirce}. Such analyses are necessary for understanding direct experiences of deprivation, but they do not fully characterize the urban contexts in which households are located. District-level analyses provide a complementary perspective by examining characteristics that may be addressed through territorial policies, such as the distribution of street markets and food-retail establishments. These contextual indicators may inform hypotheses about inequalities in the urban food environment, although ecological associations cannot be interpreted as individual-level or causal effects.

This study examines the city of São Paulo, Brazil, integrating three administrative data sources: the São Paulo Social Vulnerability Index (\textit{Índice Paulista de Vulnerabilidade Social}, IPVS), establishment records from the \textit{Relação Anual de Informações Sociais} (RAIS), and street-market information compiled by CAISAN. The data were aggregated at the district level to align their geographical resolutions. The resulting predictors describe the densities of establishments classified according to food type, the number of street markets, and the availability of establishments selling fresh or \textit{in natura} food.

The objective was to assess whether these food-environment indicators discriminate between districts classified in different IPVS groups. Conventional machine-learning classifiers were used as exploratory analytical tools, followed by an examination of the relative contribution of the predictors. Rather than treating predictive performance as evidence that food availability determines vulnerability, the analysis investigates whether publicly available food-environment variables contain information associated with the spatial distribution of social vulnerability across São Paulo's districts.

\section{Methods}

\subsection{Study design and setting}

We conducted an exploratory cross-sectional ecological study of the 96 administrative districts of the municipality of São Paulo, Brazil. The district was adopted as the unit of analysis because the available sources described social vulnerability and food-environment characteristics at different geographical resolutions. Census-sector information was therefore aggregated to the district level before integration with district-level records. The final analytical dataset included one observation per district and variables describing social vulnerability, food-retail establishments, and street markets.

\subsection{Data sources}

Three administrative data sources were integrated. Social vulnerability was characterized using the \textit{Índice Paulista de Vulnerabilidade Social} (IPVS), developed by the Fundação Sistema Estadual de Análise de Dados (SEADE). The IPVS combines socioeconomic and demographic information, including household income, education, and access to services, to classify geographical areas according to increasing levels of social vulnerability. The data used in this study refer to the 2010 edition of the index.

Information on food-retail establishments was obtained from the 2016 \textit{Relação Anual de Informações Sociais} (RAIS)\footnote{\url{https://www.rais.gov.br/}}, maintained by the Brazilian federal government. RAIS contains establishment-level administrative records, including economic activity and geographical location. The study used a filtered database containing establishments that provided food services or sold food at retail. Establishments were categorized according to the food groups available in the original database, allowing the construction of indicators related to healthy, unhealthy, fresh, and \textit{in natura} food availability.

Information on street markets (\textit{feiras livres}) was obtained from a database organized by the \textit{Câmara Interministerial de Segurança Alimentar e Nutricional} (CAISAN). These markets constitute an important component of the urban food environment because they commonly provide fruits, vegetables, and other perishable foods. The database includes the geographical distribution and number of street markets and was compiled from RAIS 2016 and the 2008--2009 Brazilian Household Budget Survey (\textit{Pesquisa de Orçamentos Familiares}, POF/IBGE). The food-environment variables describe the availability of establishments and markets within districts; they do not directly measure household food consumption, affordability, or food insecurity.

\subsection{Geographical aggregation and data integration}

The IPVS data included information for 18,363 census sectors distributed across the 96 districts of São Paulo. The number of census sectors per district ranged from 28 to 609. Because the predictors were available at different geographical resolutions, census-sector values had to be reduced to one value per district before the datasets could be integrated.

Mean- and median-based aggregation were evaluated as possible reduction procedures. To assess whether these transformations preserved the relationships present in the original data, distance matrices were calculated for the unaggregated features and for the corresponding mean- and median-aggregated features. Normalized compression distance was used as the distance measure \cite{ncd}, and agreement among the resulting matrices was assessed using Mantel tests with Pearson correlation \cite{mantel}. The comparisons were repeated using minimum, gzip, and bzip2 compression schemes.

\begin{table}[htb!]
\footnotesize
\centering
\caption{Mantel-test results comparing distance matrices obtained before and after district-level aggregation.}
\begin{tabular}{@{}llcc@{}}
\hline
\textbf{Compression} & \textbf{Comparison} & \textbf{Statistic} & \textbf{Significance} \\ \hline
min & Mean--median & 0.9219 & 0.00029997 \\ \hline
min & Mean--no reduction & 0.7770 & $9.999 \times 10^{-5}$ \\ \hline
min & Median--no reduction & 0.9114 & $9.999 \times 10^{-5}$ \\ \hline
gzip & Mean--median & 0.6425 & 0.00019998 \\ \hline
gzip & Mean--no reduction & 0.4476 & 0.0065993 \\ \hline
gzip & Median--no reduction & 0.8404 & $9.999 \times 10^{-5}$ \\ \hline
bzip2 & Mean--median & 0.8983 & $9.999 \times 10^{-5}$ \\ \hline
bzip2 & Mean--no reduction & 0.7733 & $9.999 \times 10^{-5}$ \\ \hline
bzip2 & Median--no reduction & 0.8371 & $9.999 \times 10^{-5}$ \\ \hline
\end{tabular}
\label{table:Mantel}
\end{table}

The correlations indicated substantial agreement among the representations. Mean aggregation was selected for the main analysis, resulting in a dataset with 96 district-level observations. District names and geographical identifiers were harmonized across the three databases before the variables were merged.

\subsection{Outcome definition}

The IPVS was adopted as the outcome. Its original values range from zero to seven, with higher classified values generally indicating greater social vulnerability. Table \ref{IPVS} presents the distribution of the index across the districts included in the source database.

\begin{table}[htb!]
\footnotesize
\centering
\caption{Distribution of IPVS categories across the 96 districts of São Paulo.}
\begin{tabular}{@{}c p{0.62\columnwidth} c@{}}
\hline
\textbf{Index} & \textbf{IPVS category} & \textbf{\# districts} \\ \hline
0 & Not classified & 20 \\ \hline
1 & Extremely low vulnerability & 56 \\ \hline
2 & Very low vulnerability & 7 \\ \hline
3 & Low vulnerability & 8 \\ \hline
4 & Medium vulnerability & 1 \\ \hline
5 & High vulnerability (urban) & 0 \\ \hline
6 & Very high vulnerability (urban agglomerations) & 4 \\ \hline
7 & High vulnerability (rural) & 0 \\ \hline
\end{tabular}
\label{IPVS}
\end{table}

The 20 districts assigned IPVS level zero were unclassified and were excluded because no vulnerability category was available. The analytical sample consequently comprised 76 districts. Because the remaining categories were highly imbalanced and some contained very few or no observations, the outcome was binarized. Districts classified as IPVS level 1 formed the reference class ((n=56)), whereas districts classified as levels 2--7 formed the positive class ((n=20)). These groups are referred to as `IPVS 1'' and `IPVS 2--7'', respectively, rather than as non-vulnerable and vulnerable districts, because IPVS level 2 is formally categorized as very low vulnerability.

\subsection{Food-environment predictors}

Five predictors were selected from the RAIS and CAISAN databases based on their relevance to the district food environment: density of healthy establishments, density of unhealthy establishments, number of street markets, number of natural-food establishments, and number of establishments selling \textit{in natura} food. Table \ref{table:variables} presents the variables and their observed ranges. The density indicators were retained as provided in the integrated database, whereas the remaining indicators represented district-level counts.

\begin{table}[htb!]
\footnotesize
\centering
\caption{Food-environment predictors and outcome used in the analysis.}
\begin{tabular}{@{}p{0.42\columnwidth}p{0.20\columnwidth}p{0.27\columnwidth}@{}}
\hline
\textbf{Variable} & \textbf{Type} & \textbf{Observed range} \\ \hline
Density of healthy establishments & Continuous & 8.31--192.41 \\ \hline
Density of unhealthy establishments & Continuous & 0.00--172.92 \\ \hline
Number of street markets & Integer & 0--22 \\ \hline
Number of natural-food establishments & Integer & 1--125 \\ \hline
Establishments selling \textit{in natura} food & Integer & 3--147 \\ \hline
São Paulo Social Vulnerability Index & Integer & 0--7 \\ \hline
\end{tabular}
\label{table:variables}
\end{table}

Continuous predictors were standardized using the (z)-score transformation,

\begin{equation}
x_{\mathrm{std}}=\frac{x-\mu}{\sigma},
\label{Z-Score}
\end{equation}

where $x$ is the original value and $\mu$ and $\sigma$ are the mean and standard deviation estimated from the corresponding training data. Standardization was applied within the model-fitting procedure to prevent information from held-out observations from influencing the training data.

\subsection{Classification models and evaluation}

Eight conventional classification algorithms were evaluated: K-nearest neighbors (KNN) \cite{Cover1967}, linear discriminant analysis (LDA) \cite{Fisher1936}, Naïve Bayes (NB) \cite{John1995}, decision tree (DT) \cite{Quinlan1986}, Random Forest (RF) \cite{Breiman2001}, Gradient Boosting (GB) \cite{Friedman2001}, XGBoost (XGB) \cite{Chen2016}, and support vector machine (SVM) \cite{Cortes1995}. This set included linear, probabilistic, distance-based, kernel-based, and tree-ensemble approaches, allowing the analysis to examine whether the observed discrimination depended on a particular model family.

Model performance was assessed using leave-one-out cross-validation. In each iteration, one district was retained for testing and the remaining 75 districts were used for model fitting. The procedure was repeated until every district had served once as the held-out observation. Hyperparameters were selected through grid search. All preprocessing steps that depended on the observed data, including standardization, were fitted using the training portion of each cross-validation iteration.

Because the positive IPVS 2--7 group was substantially smaller than the IPVS 1 group, the primary performance measure was the F-score:

\begin{equation}
\textit{F-score} = 
\frac{2\mathrm{TP}}
{2\mathrm{TP}+\mathrm{FP}+\mathrm{FN}},
\label{F-Score}
\end{equation}

where TP, FP, and FN denote true-positive, false-positive, and false-negative classifications, respectively. Districts in the IPVS 2--7 group were treated as positive cases. Model results were compared across the eight classifiers, and Random Forest feature importance was subsequently examined to identify the predictors contributing most strongly to classification. A permutation procedure was additionally used to assess whether the contribution of the number of natural-food establishments persisted when the predictor values were disrupted.

The analyses were implemented in Python 3.10 using Scikit-learn, NumPy, Pandas, Statistics, Matplotlib, and the XGBoost library, and were executed in Google Colaboratory.

\section{Results}

\subsection{Analytical sample and classification performance}

Of the 96 districts in the source database, 20 were excluded because they had no assigned IPVS category. The final analytical sample therefore comprised 76 districts: 56 classified as IPVS level 1 and 20 classified as IPVS levels 2--7. This distribution resulted in an imbalanced binary classification problem, with approximately 74

Table \ref{results_2} presents the mean F-scores and standard deviations obtained under the leave-one-out cross-validation procedure. Mean values ranged from 0.62 for KNN to 0.75 for XGBoost. XGBoost achieved the highest reported mean F-score, followed by LDA, Naïve Bayes, and Random Forest, with values of 0.72, 0.71, and 0.70, respectively. The remaining tree-based models, decision tree and Gradient Boosting, both obtained a mean F-score of 0.68, whereas SVM obtained 0.63.

\begin{table}[htb!]
\footnotesize
\centering
\caption{Mean F-score and standard deviation obtained by the classification models under leave-one-out cross-validation.}
\begin{tabular}{@{}lc@{}}
\hline
\textbf{Model} & \textbf{F-score} \\ \hline
KNN & 0.62 (0.49) \\ \hline
LDA & 0.72 (0.45) \\ \hline
NB & 0.71 (0.45) \\ \hline
DT & 0.68 (0.46) \\ \hline
RF & 0.70 (0.46) \\ \hline
GB & 0.68 (0.46) \\ \hline
\textbf{XGB} & \textbf{0.75 (0.43)} \\ \hline
SVM & 0.63 (0.48) \\ \hline
\end{tabular}
\label{results_2}
\end{table}

Although XGBoost produced the highest mean value, the differences among several classifiers were modest. The large standard deviations also indicate substantial variation across held-out observations. The model ranking should therefore be interpreted as exploratory rather than as evidence of a clear superiority of one algorithm. More broadly, the results indicate that the selected food-environment variables contained information that allowed the models to discriminate, to varying degrees, between the two IPVS groups.

\subsection{Relative importance of food-environment indicators}

Random Forest was used to examine the relative contribution of the predictors because its tree-based structure provides a direct measure of variable importance. This interpretive analysis was conducted separately from the comparison of predictive performance, in which XGBoost obtained the highest mean F-score.

\begin{figure}[htb!]
\centering
\includegraphics[width=1\linewidth]{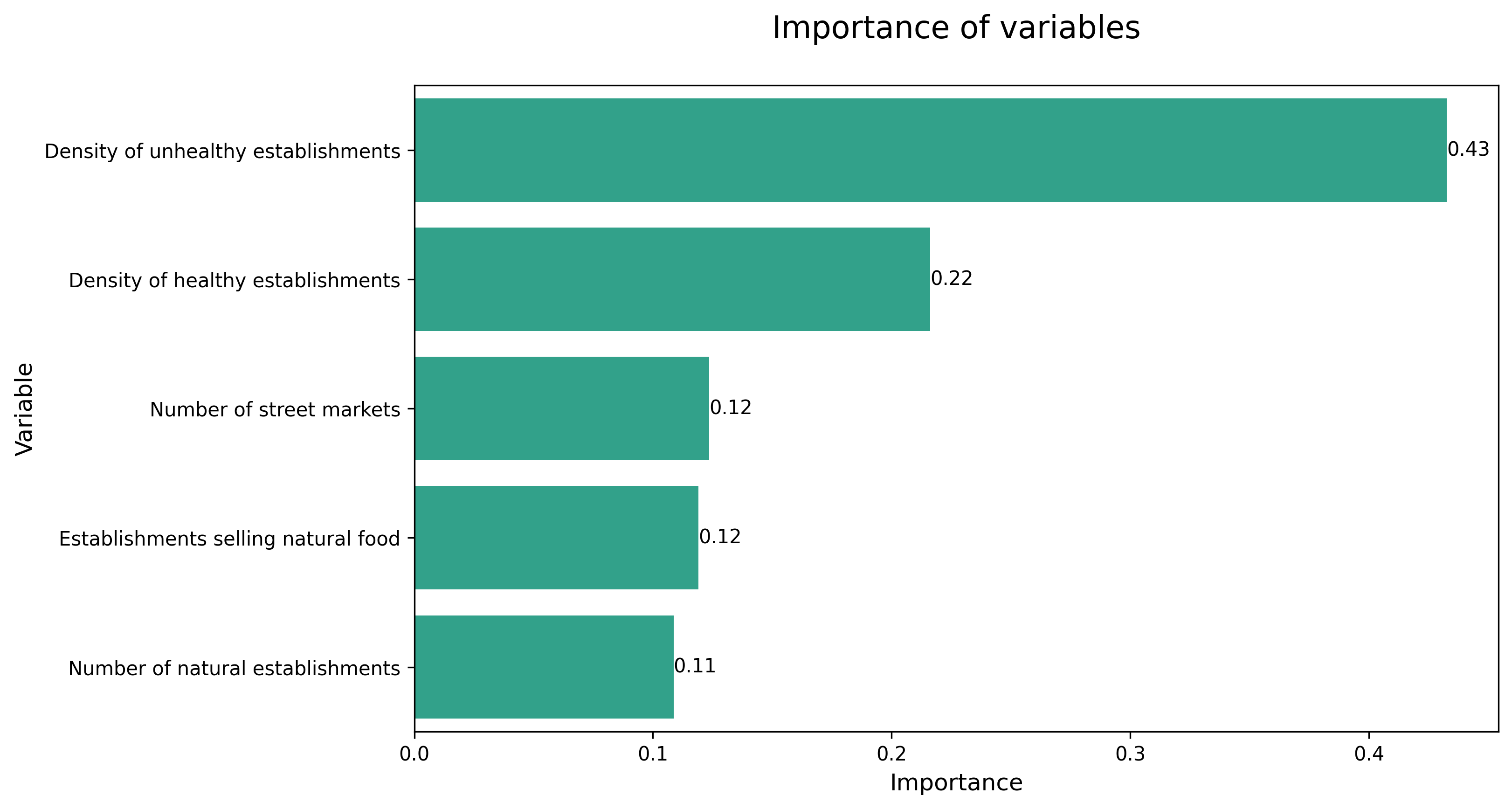}
\caption{Impurity-based importance of the food-environment variables in the Random Forest model.}
\label{fig:importances_2}
\end{figure}

Figure \ref{fig:importances_2} presents the impurity-based importance assigned to each predictor by the Random Forest model. The densities of unhealthy and healthy food establishments were the two most influential variables. Together, these indicators accounted for approximately 60\% of the model's total feature importance. The remaining importance was distributed among the number of natural-food establishments, the number of establishments selling \textit{in natura} food, and the number of street markets.

The importance values indicate how frequently and effectively the predictors contributed to splits in the fitted Random Forest. They do not represent proportions of social vulnerability explained by each variable and do not establish the direction or causal nature of the associations.

A supplementary permutation analysis was conducted to evaluate whether the contribution of the selected predictors persisted after their observed values were disrupted. Particular attention was given to the number of natural-food establishments because previous domain knowledge identified it as a potentially relevant indicator of vulnerability. This variable received non-zero permutation importance, and its contribution persisted when additional predictors were included. This result supports its relevance within the fitted classifier, although the absence of an effect estimate or uncertainty interval limits stronger inferential conclusions.

\subsection{Selected model configurations}

The hyperparameters selected for each classifier are reported in Table \ref{hyperparam_2}. These configurations were used to generate the performance estimates presented in Table \ref{results_2}.

\begin{table}[htb!]
\scriptsize
\centering
\caption{Hyperparameters used to generate the classification models.}
\begin{tabular}{@{}l p{0.73\columnwidth}@{}}
\hline
\textbf{Model} & \textbf{Hyperparameters} \\ \hline
KNN &
\texttt{leaf\_size}=1;
\texttt{n\_neighbors}=1;
\texttt{p}=2;
\texttt{weights}=\texttt{uniform}
\\ \hline

    LDA &
    \texttt{shrinkage}=\texttt{None};
    \texttt{solver}=\texttt{svd}
    \\ \hline

    NB &
    \texttt{var\_smoothing}=0.0005
    \\ \hline

    DT &
    \texttt{ccp\_alpha}=0.01;
    \texttt{criterion}=\texttt{gini};
    \texttt{max\_depth}=4;
    \texttt{max\_features}=\texttt{sqrt};
    \texttt{random\_state}=0
    \\ \hline

    RF &
    \texttt{criterion}=\texttt{entropy};
    \texttt{max\_depth}=\texttt{None};
    \texttt{max\_features}=\texttt{sqrt};
    \texttt{n\_estimators}=10;
    \texttt{random\_state}=0
    \\ \hline

    GB &
    \texttt{learning\_rate}=0.01;
    \texttt{max\_depth}=3;
    \texttt{n\_estimators}=50;
    \texttt{random\_state}=0
    \\ \hline

    XGB &
    \texttt{gamma}=1.0;
    \texttt{learning\_rate}=0.01;
    \texttt{max\_depth}=3;
    \texttt{min\_child\_weight}=0;
    \texttt{n\_estimators}=100;
    \texttt{random\_state}=0
    \\ \hline

    SVM &
    \texttt{C}=100;
    \texttt{gamma}=0.1;
    \texttt{kernel}=\texttt{rbf}
    \\ \hline
\end{tabular}
\label{hyperparam_2}

\end{table}

\section{Discussion}

This exploratory ecological analysis examined whether a small set of publicly available food-environment indicators could discriminate between district-level IPVS groups in the municipality of São Paulo. The classification models produced mean F-scores ranging from 0.62 to 0.75, with XGBoost obtaining the highest reported value. The Random Forest analysis identified the densities of healthy and unhealthy food establishments as the most influential predictors, jointly accounting for approximately 60\% of the model's total impurity-based feature importance. Taken together, the findings indicate that characteristics of the district food environment contain information associated with the spatial distribution of social vulnerability. They do not, however, establish that food-establishment availability determines vulnerability or directly represents household food insecurity.

The relevance assigned to establishment density is consistent with research showing that food environments are unevenly distributed across socioeconomic contexts \cite{walker2010disparities}. Commercial location decisions, purchasing power, transportation infrastructure, land use, and the availability of public services may jointly shape both the local food environment and broader patterns of social vulnerability. Consequently, the identified predictors may capture several interconnected aspects of urban inequality rather than an isolated effect of food-retail availability. The present findings complement household-level studies of food insecurity in Brazil \cite{Dirce} by examining contextual characteristics defined for geographical areas rather than individual households.

The distinction between contextual food availability and household food access is central to the interpretation of the results. The presence of healthy-food establishments or street markets within a district does not ensure that all residents can reach these locations or afford the products offered. Conversely, residents may purchase food outside their district or use informal food-distribution channels that are not represented in administrative records. The predictors should therefore be regarded as partial indicators of the urban food environment. They do not measure food prices, travel time, transportation barriers, product quality, household income constraints, dietary consumption, or individual experiences of food insecurity.

The similarity of the results obtained by several classifiers suggests that the observed discrimination was not restricted to a single model family. Nevertheless, XGBoost's higher mean F-score should not be interpreted as evidence of clear algorithmic superiority. The differences among the best-performing models were modest, and the large reported standard deviations indicate instability across held-out districts. The limited number of observations also increases the sensitivity of model rankings to individual districts and hyperparameter choices. The analysis is therefore more informative as an assessment of whether the selected variables contain discriminatory information than as a comparison intended to identify an optimal predictive algorithm.

The feature-importance analysis requires similar caution. Impurity-based Random Forest importance quantifies the contribution of variables to splits within the fitted trees, but it does not estimate the proportion of social vulnerability explained by each predictor. It may also favor variables with particular distributions or greater opportunities for splitting. In addition, importance values do not indicate whether a higher predictor value is associated with higher or lower vulnerability. The non-zero permutation importance observed for the number of natural-food establishments supports its contribution to classification within the fitted model, but it does not constitute evidence of an independent or causal association. Future analyses with larger samples should report out-of-sample permutation importance with uncertainty intervals and use association or dependence plots to characterize the direction of the relationships.

The study illustrates the potential of combining administrative and geospatial data to investigate urban inequality at a finer geographical scale \cite{glaeser2018big}. Similar data-driven approaches have been used to examine spatial variation in health and socioeconomic outcomes \cite{ayala2022leveraging,jean2016combining}. In the present setting, the integration of IPVS, RAIS, and CAISAN records made it possible to construct a district-level dataset without primary data collection. This approach may support the initial identification of geographical patterns requiring closer investigation. It should not replace household surveys, field assessments, or consultation with local communities when defining areas for intervention.

Several limitations constrain the conclusions. First, the ecological design precludes inference about individual households and introduces the possibility of ecological fallacy. Second, the analytical sample contained only 76 districts, including 20 in the IPVS 2--7 group. This small and imbalanced sample limits model stability and the precision of performance estimates. Third, binarizing the IPVS was necessary because some categories contained very few observations, but it combined districts with substantively different vulnerability levels and placed the formally ``very low vulnerability'' IPVS level 2 in the same group as higher levels. Fourth, 20 unclassified districts were excluded, and the available data do not establish whether their exclusion introduced systematic selection bias.

Additional limitations arise from the integration of administrative sources. The IPVS and food-environment indicators refer to different collection periods, which may weaken their temporal correspondence. Governmental databases are also updated at intervals of several years, making external temporal validation difficult. Establishments absent from formal administrative records, changes in business activity, and errors in geographical classification may not be captured. The aggregation of census-sector information to districts further masks within-district heterogeneity, particularly in geographically large or socioeconomically diverse districts. Although the aggregation analysis indicated substantial agreement among the evaluated representations, district-level averages cannot preserve all local variation.

The evaluation procedure also warrants caution. Leave-one-out cross-validation maximized the amount of training data in each iteration, but each test partition contained only one district. Fold-level F-scores derived from individual observations are difficult to interpret, and the resulting averages and standard deviations may not provide a stable estimate of population-level performance. Moreover, unless hyperparameter selection is conducted entirely within each training partition, grid search may introduce optimistic bias. A future confirmatory analysis should generate pooled out-of-fold predictions, report a single confusion matrix and class-specific metrics, compare the models with majority-class and simple statistical baselines, and employ nested repeated cross-validation where feasible.

Despite these limitations, the study identifies a reproducible direction for investigating the relationship between food environments and social vulnerability using Brazilian public data. The results support further examination of establishment density, street markets, and fresh-food availability as contextual indicators, rather than demonstrating that changes in these variables would directly reduce vulnerability. Future work could incorporate food prices, accessibility through public transportation, population-adjusted exposure, spatial autocorrelation, within-district heterogeneity, and longitudinal changes in the urban food environment. Extending the analysis to additional municipalities could increase the sample size, but such comparisons would need to account explicitly for heterogeneity among cities and differences in administrative data coverage.

From a policy perspective, the findings may help identify indicators that deserve consideration in more comprehensive territorial assessments. Decisions concerning food-access interventions should combine administrative evidence with direct measures of affordability, mobility, household food insecurity, and local demand. Under this interpretation, machine-learning models serve as exploratory tools for detecting patterns and prioritizing further investigation, rather than as definitive instruments for classifying districts or prescribing interventions.

\section{Conclusion}

This exploratory ecological study found that publicly available indicators of the district food environment contained information associated with the distribution of social vulnerability across São Paulo. Among the evaluated variables, the densities of healthy and unhealthy food establishments contributed most strongly to the Random Forest classification, while XGBoost obtained the highest reported mean F-score among the eight classifiers. These results suggest that the local composition of food-retail establishments may help characterize territorial differences represented by the IPVS.

The findings should not be interpreted as evidence that food-establishment availability causes social vulnerability or directly measures household food insecurity. The small and imbalanced sample, the aggregation of heterogeneous census sectors into districts, the temporal differences among data sources, and the binarization of the IPVS restrict the generalizability of the analysis. The models are therefore better understood as exploratory instruments for identifying contextual patterns than as tools for definitive district classification or policy prescription.

Even with these limitations, the integration of IPVS, RAIS, and CAISAN records demonstrates the value of Brazilian administrative data for examining urban food environments at a submunicipal scale. Future studies should combine these indicators with measures of food affordability, mobility, household food insecurity, population exposure, and spatial dependence. Such evidence would allow a more complete assessment of whether and how inequalities in the urban food environment contribute to broader patterns of socioeconomic disadvantage.

\section*{Acknowledgments}
The authors of this work would like to thank the support from the São Paulo Research Foundation (FAPESP grants \verb|#2019/07665-4|, \verb|#2020/16578-5|,\verb|2024/06402-8|, \verb|#2024/16994-0|),
Fundação de Apoio à Universidade de São Paulo (FUSP) - Project no 3541, the IBM Corporation, the CEPID-CeMEAI/ICMC-USP (CEPID, FAPESP grant \verb|#2013/07375-0|), CNPq (grants \verb|#307085/2018-0|, \verb|#406774/2022-6|, \verb|#385292/2025-2|); and CAPES.

\bibliographystyle{IEEEtran}
\bibliography{refs}
\end{document}